\documentclass[conference]{IEEEtran}
\IEEEoverridecommandlockouts

\usepackage{cite}
\usepackage{amsmath,amssymb,amsfonts}
\usepackage{algorithmic}
\usepackage{graphicx}
\usepackage{textcomp}
\usepackage{xcolor}
\def\BibTeX{{\rm B\kern-.05em{\sc i\kern-.025em b}\kern-.08em
    T\kern-.1667em\lower.7ex\hbox{E}\kern-.125emX}}

\usepackage{booktabs}
\usepackage{listings}
\usepackage{tcolorbox}
\usepackage{url}
\usepackage{comment}

\def\BibTeX{{\rm B\kern-.05em{\sc i\kern-.025em b}\kern-.08em
    T\kern-.1667em\lower.7ex\hbox{E}\kern-.125emX}}

\providecommand{\Description}[1]{}

\providecommand{\Description}[1]{}

\newcommand{\kw}[1]{\textcolor{black}{\textbf{#1}}}
\newcommand{\akw}[1]{\textcolor{black}{\textbf{#1}}}
\newcommand{\iden}[1]{\textcolor{teal}{#1}}

\newtcolorbox{syntaxbox}{
  colback=gray!8,
  colframe=gray!40,
  boxrule=0.4pt,
  arc=2pt,
  left=6pt,
  right=6pt,
  top=4pt,
  bottom=4pt
}
\begin{document}

\title{TLSQL: Table Learning Structured Query Language

}

\author{
\IEEEauthorblockN{
Feiyang Chen\textsuperscript{1},
Ken Zhong\textsuperscript{1},
Aoqian Zhang\textsuperscript{2},
Zheng Wang\textsuperscript{1},
Li Pan\textsuperscript{1},
Jianhua Li\textsuperscript{1}
}
\IEEEauthorblockA{
\textsuperscript{1}\textit{Shanghai Jiao Tong University}, Shanghai, China.
\textsuperscript{2}\textit{Beijing Institute of Technology}, Beijing, China\\
\texttt{\{feiyang73, zhongken, wzheng, panli, lijh888\}@sjtu.edu.cn}; \texttt{aoqian.zhang@bit.edu.cn}
}
}


\maketitle

\begin{abstract}
Table learning has recently emerged as an important paradigm at the intersection of database systems and machine learning.
However, applying table learning in practice often requires exporting data from databases and building complex external machine learning pipelines, which disrupts the SQL-centric workflow commonly used by database practitioners.
We present \textbf{TLSQL} (\textbf{T}able \textbf{L}earning \textbf{S}tructured \textbf{Q}uery \textbf{L}anguage), a lightweight SQL-like interface for specifying table learning tasks over SQL-centric data systems.
TLSQL allows users to declaratively define predictive tasks using three simple constructs: \textbf{PREDICT VALUE}, \textbf{TRAIN WITH}, and \textbf{VALIDATE WITH}. 
TLSQL specifications are compiled into standard SQL queries executed by the database engine and task descriptions consumed by downstream table learning frameworks.
This design allows users to focus on modeling rather than low-level data preparation and pipeline orchestration.
Our demonstration shows that TLSQL enables end-to-end multi-table learning workflows while preserving familiar SQL-based data processing environments.
Our code is available at \url{https://github.com/tlsql-project/tlsql}.
\end{abstract}


\section{\textbf{Introduction}}
The rapid adoption of artificial intelligence has brought table learning~\cite{borisov2022deep} to the forefront of research at the intersection of database systems and machine learning. In particular, recent advances in relational table learning (RTL)~\cite{zahradnik2023deep,robinson2024relbench} enable end-to-end machine learning directly over relational databases. Consequently, table learning, especially in multi-table settings, is increasingly recognized as a promising paradigm for integrating machine learning with modern database systems.

Despite this progress, applying table learning in practice remains challenging for database users. Existing solutions typically follow a database-decoupled workflow: data must be exported from databases, manually preprocessed, and then consumed by programmatic machine learning pipelines. This process increases system complexity and breaks the SQL-centric workflow that database practitioners rely on.

SQL-like abstractions have proven effective in simplifying large-scale data processing, as demonstrated by systems such as Spark SQL~\cite{armbrust2015spark} and Hive~\cite{Thusoo2009Hive}. However, existing SQL-based machine learning interfaces remain limited. Many systems are tightly coupled with specific platforms and rely on built-in learning components, restricting portability across heterogeneous databases. In addition, most approaches focus on single-table learning and provide complex SQL extensions that increase the cognitive burden for users. These limitations hinder the integration of machine learning into database-centric workflows.

To address these challenges, we propose \textbf{TLSQL} (\textbf{T}able \textbf{L}earning \textbf{S}tructured \textbf{Q}uery \textbf{L}anguage), a lightweight SQL-like interface for specifying table learning tasks over SQL-centric data systems.
TLSQL is designed with three key properties. First, it is database-agnostic: TLSQL specifications are compiled into standard SQL queries that can be executed on arbitrary SQL-compatible data systems. Second, TLSQL naturally supports multi-table learning by allowing users to declaratively specify data from multiple tables within a single learning task. Third, TLSQL adopts a simplified, task-oriented design that avoids the full complexity of SQL while preserving the expressiveness required for table learning workflows.

TLSQL is implemented as a lightweight Python library that compiles SQL-like specifications into standard SQL queries and structured learning task descriptions. The generated SQL queries are executed by the underlying database engine, while the task descriptions are consumed by downstream table learning frameworks. This design allows users to focus on modeling and analysis, rather than low-level data preparation and pipeline management.
Specifically, TLSQL enables users to declaratively define machine learning tasks through three core constructs:
\begin{itemize}
  \item \texttt{PREDICT VALUE}, which declares the prediction target and task type (e.g., classification or regression).
  \item \texttt{TRAIN WITH}, which specifies the tables, columns, and filtering predicates used to construct the training dataset.
  \item \texttt{VALIDATE WITH}, which defines validation data for model selection and hyperparameter tuning.
\end{itemize}

In summary, TLSQL bridges SQL-based data systems and modern table learning frameworks through a concise declarative interface. By preserving the familiar SQL-centric workflow while enabling multi-table learning tasks, TLSQL significantly lowers the barrier for database users to integrate machine learning into data-centric environments.

\section{\textbf{System Overview}}

\subsection{\textbf{Language Syntax}}
To facilitate table-centric machine learning directly over relational databases, TLSQL introduces three core statements: \texttt{PREDICT}, \texttt{TRAIN}, and \texttt{VALIDATE}.
These statements declaratively specify the prediction task, the training dataset, and the validation dataset.

\subsubsection{\textbf{PREDICT Statement}}
The \texttt{PREDICT} statement defines the prediction target, the task type (classification or regression), and optional filtering conditions for the test set. Its syntax is as follows:

\begin{syntaxbox}
\small
\texttt{\noindent
\kw{PREDICT VALUE} (\iden{column\_selector}, \iden{TASK\_TYPE})\\[0.3em]
\akw{FROM} \iden{table}\\[0.3em]
{[}\akw{WHERE} \iden{conditions}{]}
}
\end{syntaxbox}

Here, \texttt{column\_selector} can be either \texttt{table.column} or \texttt{column}; \texttt{TASK\_TYPE} must be \texttt{CLF} (classification) or \texttt{REG} (regression). The \texttt{FROM} clause specifies the table containing the prediction target, and the optional \texttt{WHERE} clause allows row-level filtering in the test set.

\subsubsection{\textbf{TRAIN Statement}}
The \texttt{TRAIN} statement specifies the provenance of the training dataset. Its syntax is:

\begin{syntaxbox}
\small
\texttt{\noindent
\kw{TRAIN WITH} \iden{column\_selector}\\[0.3em]
\akw{FROM} \iden{table1, table2, ...}\\[0.3em]
{[}\akw{WHERE} \iden{conditions}{]}
}
\end{syntaxbox}

The \texttt{FROM} clause may list one or more tables, separated by commas. The optional \texttt{WHERE} clause supports Boolean expressions across multiple tables. During execution, TLSQL groups column selectors by table and routes cross-table predicates to their corresponding tables, generating independent standard SQL queries for each table.

\subsubsection{\textbf{VALIDATE Statement}}
The \texttt{VALIDATE} statement specifies the validation dataset.
Since general machine learning methods require the validation and test sets to be semantically consistent (i.e., drawn from the same sample space and label space), the syntax of \texttt{VALIDATE} mirrors that of \texttt{PREDICT}, differing only in the leading keyword:

\begin{syntaxbox}
\small
\texttt{\noindent
\kw{VALIDATE WITH} \iden{column\_selector} \\[0.3em]
\akw{FROM} \iden{table}\\[0.3em]
{[}\akw{WHERE} \iden{conditions}{]}
}
\end{syntaxbox}

In this statement, the task type is omitted and implicitly inherited from the corresponding \texttt{PREDICT} specification.

\subsection{\textbf{Compilation Pipeline}}
\begin{figure}[!t]
  \centering
  \includegraphics[width=\linewidth]{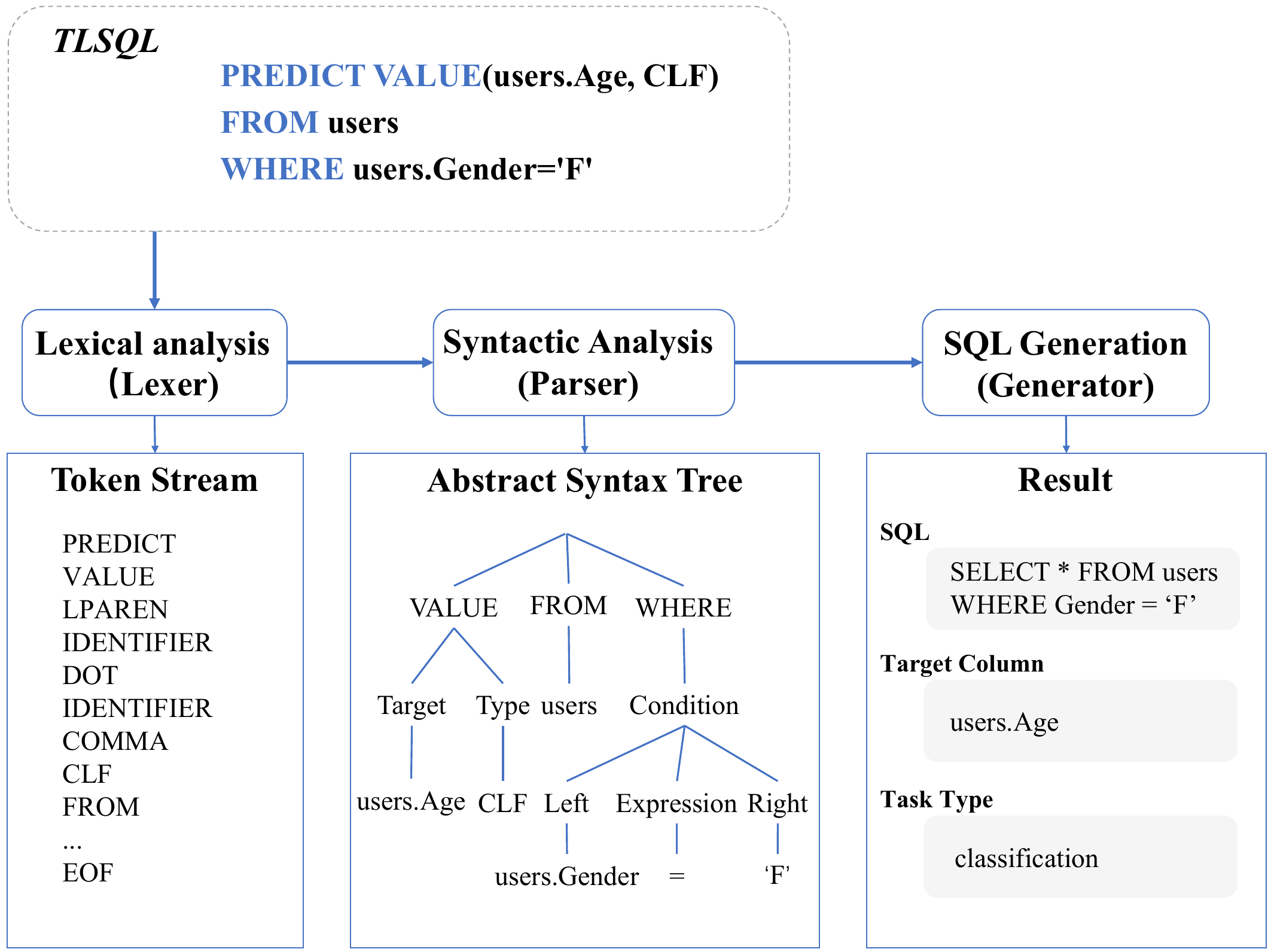}
  \caption{TLSQL compilation pipeline illustrated using the TML1M dataset~\cite{li2024rllm}, which comprises three interconnected tables: \texttt{users}, \texttt{movies}, and \texttt{ratings}.}
  \label{fig:workflow}
\end{figure}

As shown in Figure~\ref{fig:workflow}, TLSQL follows a three-stage processing pipeline consisting of a \textbf{Lexer}, a \textbf{Parser}, and a \textbf{SQLGenerator}.
This modular architecture preserves syntactic compatibility with SQL while supporting multi-table specifications and robust data partitioning logic.

\subsubsection{\textbf{Lexer}}
The Lexer performs lexical analysis by identifying keywords (e.g., \texttt{PREDICT}, \texttt{TRAIN}, \texttt{VALIDATE}, \texttt{FROM}, \texttt{WHERE}), operators, and identifiers, encapsulating them into \texttt{Token} objects annotated with positional metadata. Invalid characters or unclosed strings trigger precise, position-aware exceptions, ensuring a structurally coherent token stream.

\subsubsection{\textbf{Parser}}
The Parser executes syntactic analysis by consuming the token stream and applying the TLSQL grammar. Using recursive descent, it delineates clause boundaries, resolves nested structures, validates composition, and constructs a hierarchical AST. The parser also detects and localizes syntactic errors, providing detailed diagnostics. This AST captures the logical structure of TLSQL statements and forms the basis for subsequent SQL generation.

\begin{figure*}[!t]
  \centering
  \includegraphics[width=1.0\linewidth]{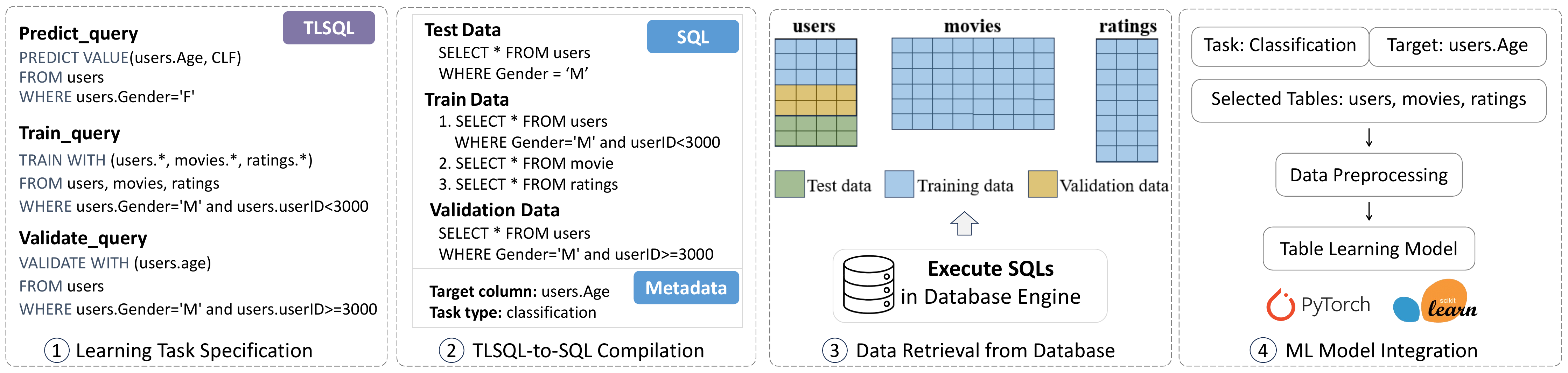}
  \caption{End-to-end table learning with TLSQL.}
  \label{fig:rtl_demo}
   \Description{1}
\end{figure*}

\subsubsection{\textbf{SQLGenerator}}
The SQLGenerator traverses the AST to produce standard SQL queries and extract structured table learning task descriptions. Specifically, it:
\begin{itemize}
    \item Groups \texttt{WITH} columns by table.
    \item Decomposes \texttt{WHERE} conditions into atomic predicates and assigns them to the corresponding tables.
    \item Extracts the prediction target column and task type from the \texttt{VALUE} clause.
\end{itemize}

The generated SQL can be executed directly by existing database systems, while the structured task descriptions are consumed by downstream table learning frameworks. 
This enables a single TLSQL program to orchestrate an end-to-end workflow encompassing data extraction, task specification, and model training.

\section{\textbf{Demonstration}}

\subsection{\textbf{End-to-End Table Learning Workflow}}

Figure~\ref{fig:rtl_demo} illustrates an end-to-end table learning workflow enabled by TLSQL, demonstrating its integration with the RTL method over relational databases. In this demonstration, we continue to use the TML1M dataset stored in a MySQL database. The workflow proceeds through four steps, as annotated in the figure:

\textcircled{1}~\textbf{Learning Task Specification.}
Users declare a relational learning task using TLSQL statements in a SQL-like form.
In this example, the \texttt{PREDICT} clause specifies a classification task on the \texttt{Age} attribute of female users.
The \texttt{TRAIN} and \texttt{VALIDATE} clauses define the relational tables and conditions used to construct the training and validation datasets.

\textcircled{2}~\textbf{TLSQL-to-SQL Compilation.}
TLSQL translates the declarative task specification into standard SQL queries together with structured task metadata (e.g., prediction target and task type), forming an executable plan for downstream processing.

\textcircled{3}~\textbf{Data Retrieval from Database.}
The generated SQL queries are executed by the database engine to retrieve the required tables for training, validation, and prediction. The retrieved data preserves the original table structures without additional preprocessing or manual feature engineering.

\textcircled{4}~\textbf{ML Model Integration.}
The retrieved tables and task metadata are passed to the learning pipeline and transformed into model-ready representations. These inputs can then be consumed by downstream relational learning frameworks (e.g., BRIDGE) implemented in systems such as PyTorch or scikit-learn.

\begin{figure}[!t]
  \centering
\begin{lstlisting}[language=Python, basicstyle=\ttfamily\small, morekeywords={@classmethod}]
@classmethod
def convert(tlsql: str) -> ConversionResult:
    # Parse TLSQL to AST
    parser = Parser(tlsql)
    ast = parser.parse()
    
    # Generate SQL with metadata
    generator = SQLGenerator()
    return generator.build(ast)

if __name__ == "__main__":
    predict_query = """
    PREDICT VALUE(users.Age, CLF) 
    FROM users 
    WHERE users.Gender='F'"""
    result = convert(predict_query)
\end{lstlisting}
  \caption{A minimal example of the TLSQL programming interface.}
  \label{fig:usage}
\end{figure}

\subsection{\textbf{Programming Interface}}
Figure~\ref{fig:usage} illustrates the core implementation of the TLSQL conversion system.
The \texttt{convert} function translates a TLSQL query into executable SQL statements while simultaneously extracting structured metadata required for downstream learning tasks.
Internally, the conversion follows a two-stage pipeline consisting of parsing and generation.
By encapsulating the entire workflow within a single function call, TLSQL provides a concise and user-friendly programming interface.

The returned \texttt{ConversionResult} encapsulates the SQL queries generated for each
referenced table as well as structured metadata describing the learning task.
This abstraction enables users to directly submit the generated SQL queries to the
underlying database system, while passing the metadata to downstream components such as
feature preprocessors, data loaders, or table learning models.
This abstraction allows users to focus on high-level task specification rather than low-level data extraction or pipeline orchestration.

\begin{figure}[!t]
  \centering
  \includegraphics[width=0.9\linewidth]{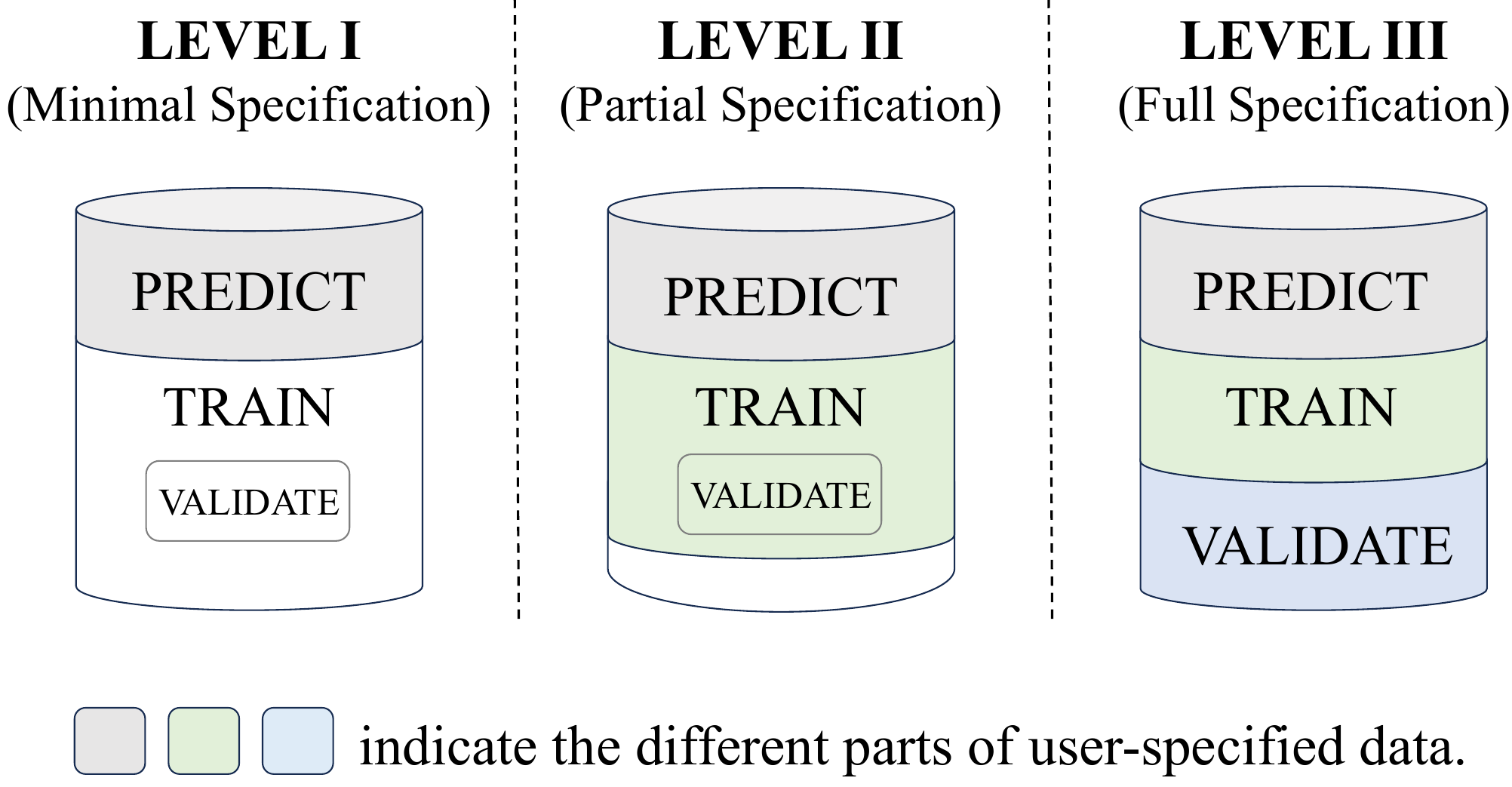}
  \caption{Three-level specification in TLSQL.}
  \label{fig:tlsql_level}
   \Description{1}
\end{figure}

\subsection{\textbf{Three-Level Task Specification}}
Figure~\ref{fig:tlsql_level} illustrates the three-level task specification mechanism in TLSQL. The \texttt{convert\_workflow} function accepts multiple TLSQL queries that jointly define a learning workflow. TLSQL analyzes the relationships among these queries and consolidates them into a unified execution plan. The system supports three specification levels with increasing user control:

\begin{itemize}
    \item \textbf{Level I (Minimal Specification).}  
    Users specify only the \texttt{PREDICT} statement. If the \texttt{TRAIN} statement is omitted, TLSQL automatically uses all remaining data as training data. When the \texttt{VALIDATE} statement is not specified, the system applies default validation (e.g., 5-fold cross-validation or temporal splits).

    \item \textbf{Level II (Partial Specification).}  
    Users specify both the \texttt{PREDICT} and \texttt{TRAIN} statements, while the \texttt{VALIDATE} statement remains optional. In this case, TLSQL applies the same default validation strategy as in Level I.

    \item \textbf{Level III (Full Specification).}  
    Users explicitly define all three statements—\texttt{PREDICT}, \texttt{TRAIN}, and \texttt{VALIDATE}—to gain full control over data selection, partitioning, and evaluation strategies.
\end{itemize}

This design allows users to start with minimal input and add more control when needed, making TLSQL easy to use for beginners while remaining flexible for advanced users.

\begin{table}[!t]
\centering
\begin{tabular}{lcc}
\toprule
\textbf{Property} & \textbf{SQL-based ML Systems} & \textbf{TLSQL} \\
\midrule
ML interface & Function APIs & \textbf{Declarative clauses} \\
Model flexibility & Limited (built-in) & \textbf{Flexible (external)} \\
Abstraction level & Tuple / feature & \textbf{Table-level} \\
Multi-table learning & Limited & \textbf{Native support} \\
Portability & Platform-specific & \textbf{DB-agnostic} \\
Ease of use & Complex & \textbf{Simple} \\
\bottomrule
\end{tabular}
\vspace{1pt}
\caption{Comparison between SQL-based ML systems (e.g., PostgresML, MADLib, EvaDB, GaussML) and TLSQL.}
\label{tab:comparison}
\end{table}

\section{\textbf{Discussion}}

\subsection{\textbf{Understanding TLSQL Through SQL}}

TLSQL adopts a minimalist, task-oriented design that introduces three high-level statements while omitting SQL features unnecessary for learning task specification. The goal is to support end-to-end table learning, where selecting relevant relational data is more important than complex data manipulation.

First, TLSQL naturally supports multi-table learning. While SQL queries typically produce a single output relation, modern relational learning methods operate over collections of interrelated tables representing different entities. TLSQL allows users to specify multiple related tables directly, aligning with the data requirements of contemporary table learning frameworks.

Second, TLSQL simplifies join logic to improve usability and maintain a separation between data specification and model reasoning. Nested queries are disallowed, and \texttt{WHERE} clauses are restricted to simple Boolean predicates. Explicit cross-table joins are omitted and handled by downstream learning models, enabling users to work with independent tables without constructing complex join pipelines. This loosely coupled design is particularly suitable for data lake environments, where relationships among tables may be implicit or weakly defined~\cite{pan2026lakemlb}.

Third, TLSQL omits traditional data transformation operators such as aggregations (e.g., \texttt{SUM} and \texttt{AVG}). Modern table learning models can directly consume relational data and learn useful representations automatically, reducing the need for manual feature engineering.

Overall, TLSQL focuses on providing a simple and lightweight interface for specifying table learning tasks rather than replacing SQL or serving as a general-purpose ML language. Future extensions may introduce additional capabilities, but the core principle of maintaining a minimal and declarative interface will remain unchanged.

\subsection{\textbf{TLSQL vs. SQL-based ML Systems}}

Existing SQL-based ML systems, such as PostgresML, MADLib, EvaDB, and GaussML, integrate machine learning capabilities into database systems by supporting model training and inference through SQL interfaces. These systems primarily focus on ML execution within the database engine, often exposing built-in models or function-based APIs. In contrast, TLSQL targets a different abstraction level: declaratively specifying end-to-end table learning workflows over relational data.

As summarized in Table~\ref{tab:comparison}, SQL-based ML systems typically require users to manage feature construction, data preparation, and model invocation through APIs. TLSQL instead introduces declarative clauses, including \texttt{PREDICT VALUE}, \texttt{TRAIN WITH}, and \texttt{VALIDATE WITH}, enabling users to specify predictive tasks directly at the table level. This abstraction naturally supports multi-table learning scenarios involving multiple related relational entities.

TLSQL further separates task specification from model execution by compiling task descriptions into SQL queries and learning specifications for external table learning frameworks. This design provides greater model flexibility and portability across databases and learning backends. While SQL-based ML systems primarily extend databases with ML execution capabilities, TLSQL provides a declarative layer for specifying and orchestrating table learning workflows.

\section{\textbf{Conclusion}}
In this work, we presented \textbf{TLSQL}, a declarative SQL-like interface for table learning over modern relational databases.
TLSQL allows users to specify predictive tasks using simple constructs, \texttt{PREDICT VALUE}, \texttt{TRAIN WITH}, and \texttt{VALIDATE WITH}, eliminating the need for complex manual data export, feature engineering, or low-level pipeline management. 
We believe TLSQL provides a strong foundation for future research in database-centric learning, offering both a conceptual framework and a practical prototype.
In future work, we will continue evolving TLSQL to better support emerging machine learning applications and advanced learning models in SQL-centric data management systems.

\bibliographystyle{IEEEtran}
\bibliography{sample}
\end{document}